\documentclass[12pt]{article}
\usepackage{amsmath}
\usepackage{graphicx}
\usepackage{natbib} 
\usepackage{url} 
\usepackage{amsbsy}
\usepackage{bm}
\usepackage[title]{appendix}

\newcommand{\blind}{0}

\begin{document}

\bibliographystyle{Chicago}

\def\spacingset#1{\renewcommand{\baselinestretch}%
{#1}\small\normalsize} \spacingset{1}


\if0\blind
{
  \title{\bf Predicting the Output From a Stochastic Computer Model When a Deterministic Approximation is Available}
  \author{Evan Baker \thanks{
    The authors gratefully acknowledge funding provided by the Engineering and Physical Sciences Research Council}\hspace{.2cm}\\
    Department of Mathematics, University of Exeter, UK
    and \\
    Peter Challenor \\
    Department of Mathematics, University of Exeter, UK \\
    and \\
    Matt Eames \\
    Department of Engineering, University of Exeter, UK}
  \maketitle
} \fi

\if1\blind
{
  \bigskip
  \bigskip
  \bigskip
  \begin{center}
    {\LARGE\bf Predicting the Output From a Stochastic Computer Model When a Deterministic Approximation is Available}
\end{center}
  \medskip
} \fi

\bigskip
\begin{abstract}
The analysis of computer models can be aided by the construction of surrogate models, or emulators, that statistically model the numerical computer model. Increasingly, computer models are becoming stochastic, yielding different outputs each time they are run, even if the same input values are used. Stochastic computer models are more difficult to analyse and more difficult to emulate - often requiring substantially more computer model runs to fit. We present a method of using deterministic approximations of the computer model to better construct an emulator. The method is applied to numerous toy examples, as well as an idealistic epidemiology model, and a model from the building performance field.
\end{abstract}

\noindent%
{\it Keywords: Gaussian process; Heteroscedastic; Emulator; Design; Simulator} 
\vfill

\newpage
\spacingset{1} 

\section{Introduction} \label{sec:Introduction}
Complex real world systems can be modelled using computer simulators, allowing experimentation to be conducted where physical experiments may be too costly or infeasible. Typically such computer simulators are deterministic, yielding the same output every time the simulator is run if the same input parameter values are used. Increasingly however, simulators are becoming stochastic, including some random component in their code to account for perceived `randomness' in a system. Running such simulators with the same input parameter values does not yield the same output value each time. Instead there is some intrinsic variance in the computer model output.

Often simulators are computationally expensive, and so to facilitate analysis, a statistical approximation of these models is used. Known as emulators, these use previously obtained values from the simulator to obtain fast predictions for new outputs of the simulator.

For deterministic models, the Gaussian process emulator is a common choice of emulator \citep{OHagan2006}. The principal idea is that the outputs from the computer model $\eta(\cdot)$ are  treated as a realisation from a Gaussian process with a prior mean function $m(\cdot)$ and a prior covariance function $K(\cdot, \cdot)$. Predictions can then be obtained by conditioning on data $\textbf{y} = \eta(X)$ \citep{Rasmussen2004}.

This framework can be extended to model stochastic simulators by the addition of an independent noise function with variance $\delta(\cdot)$. If the simulator is believed to be homoscedastic, this can simply be a constant. Alternatively, if the homoscedastic assumption is too strong, one can also model the (log) variance with another Gaussian process, with mean function $m_{\delta}(\cdot)$ and covariance function $K_{\delta}(\cdot, \cdot)$ \citep{Goldberg1997, Boukouvalas2009, Binois2016}. The log variance is modelled rather than simply the variance to constrain predicted values of the variance to be greater than zero.

Such a model is then very flexible, with a non-parametric form for both the mean and variance, allowing for many different stochastic simulators to be emulated. The downside to this flexibility is that far more data is required to properly estimate the form of the mean (and variance). A rule of thumb for deterministic emulators is that at least 10 data points per input dimensions are required to fit an emulator \citep{Loeppky2009}, whereas \cite{Binois2017} use 50 times this number when comparing different methods of choosing data point locations on a 1D toy stochastic simulator.

A larger required amount of data is to be expected for more complicated simulators (and stochastic simulators are certainty a more complex class of simulator), but such a high number of required runs can be prohibitive. In this article we attempt to alleviate this problem by leveraging a unique tool that computer simulators provide over physical experiments:  often deterministic approximations are available.
Section \ref{sec:Intuition} will provide the motivation for the method. Section \ref{sec:DetHetGP} will then formally present this method. Section \ref{sec:Performance} discusses the advantages and disadvantages this model provides, using numerous simulation experiments on toy simulators. Section \ref{sec:Examples} then applies this method to a simple agent based model and a stochastic building performance simulator. Concluding remarks are given in section \ref{sec:Conclusions}.

\section{Intuition} \label{sec:Intuition}

This section aims to build up the intuition behind the new model in a natural way, starting from the heteroscedastic Gaussian process emulator mentioned in section \ref{sec:Introduction}.\\
In this article, the mean functions $m(\textbf{x})$ and $m_\delta(\textbf{x})$ are taken to be the linear functions  $\beta_0$ + $\textbf{x}^T\bm{\beta}$ and $\beta_{\delta_0}$ + $\textbf{x}^T\bm{\beta}_\delta$ respectively. Similarly, the covariance functions $K(\textbf{x}, \textbf{x}')$ and $K_{\delta}(\textbf{x}, \textbf{x}')$ are respectively taken to be $\alpha^2\prod_{i=1}^{d}\exp(-(\frac{x_i - x_i'}{l_i})^2)$ and $\alpha_\delta^2\prod_{i=1}^{d}\exp(-(\frac{x_i - x_i'}{l_{\delta i}})^2)$, where $d$ is the dimension of $\textbf{x}$ (these are the squared exponential correlation functions, multiplied by a variance parameter $\alpha^2$).
Additionally, in this article, parameters are taken fixed as their posterior modes, estimated via the optimizing function in Stan \citep{Team2015}.
The priors for the $\beta_0$, $\bm{\beta}$, $\beta_{\delta_0}$ and $\bm{\beta}_\delta$ will be $N(0,10)$; $\alpha$ and $\alpha_\delta$ will have $Inverse-Gamma(2,1)$ priors; and the length scales $l_i$ and $l_{\delta i}$ will have $Gamma(4,4)$ priors.

\begin{figure}[!ht]
\includegraphics[width=\textwidth]{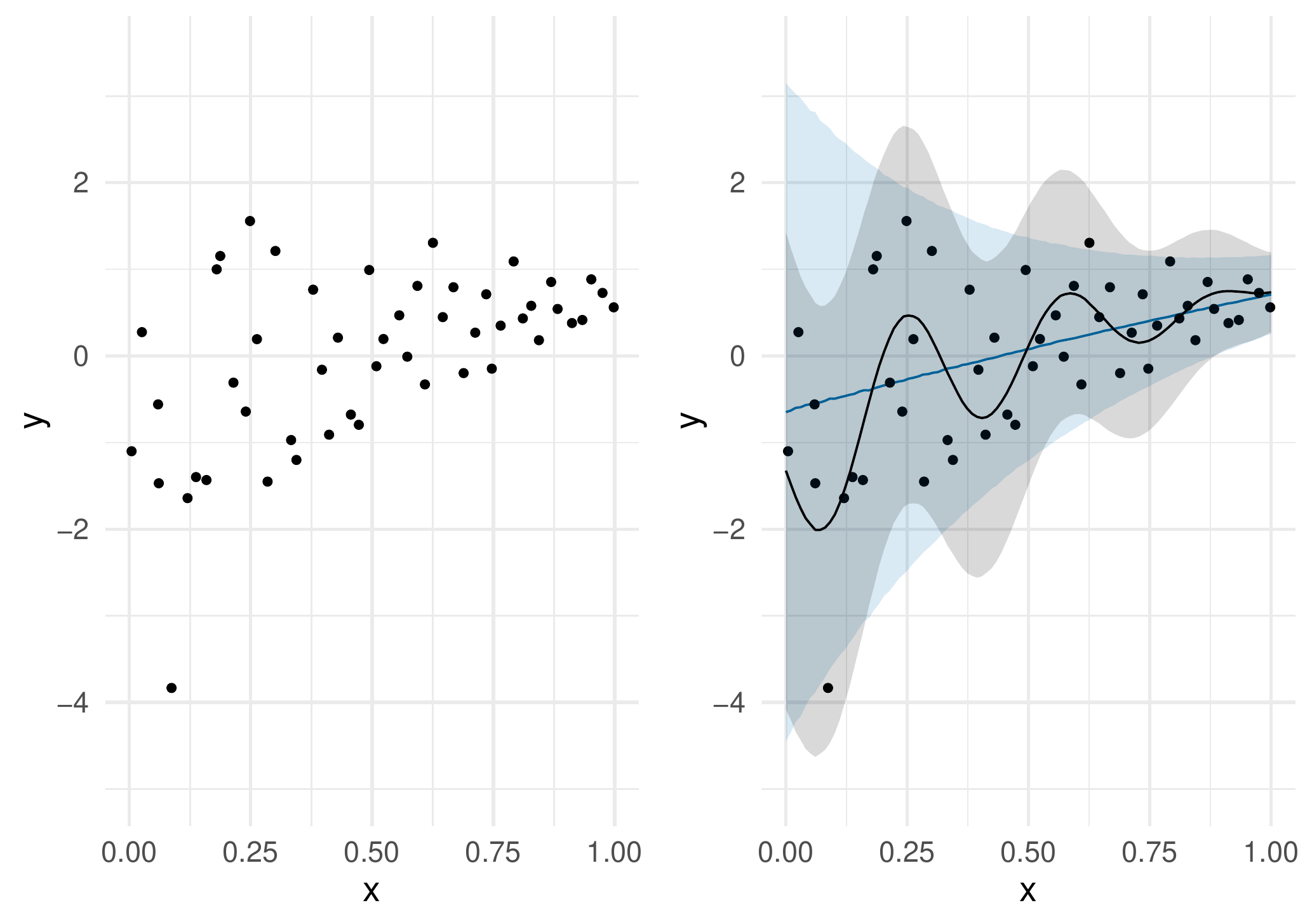}
\caption{50 Evaluations of the toy simulator from equation \ref{eq:toy} (left), and the respective heteroscedastic emulator predictions (right). The True mean and $95\%$ interval are superimposed in black, and the emulator mean and $95\%$ interval are in blue}
\label{fig:HetGPplot}
\end{figure}

To begin with, we will look at the toy stochastic simulator given in equation \ref{eq:toy}.

\begin{equation} \label{eq:toy}
\begin{gathered}
	\eta(x) = (1-x)sin(\pi + 6\pi x) + log(0.2+x) + (1.2-x)\epsilon\\
	\epsilon \sim N(0, 1)
\end{gathered}
\end{equation}
 Evaluating this toy simulator on 50 points sampled from $[0,1]$ using a maximin latin hypercube \citep{McKay2000}, and standardising the data, we obtain the plot in left of figure \ref{fig:HetGPplot}.
 
With only the plot on the left of figure \ref{fig:HetGPplot}, the challenge of flexibly modelling a heteroscedastic process becomes clearer. Perceived trends can be `true', or they can also just be artefacts of the stochasticity. This makes it difficult to discern the correct shapes for the mean and variance functions of a simulator without a large number of data points. The plot on the right of figure \ref{fig:HetGPplot} shows the predictions for the mean and the $95\%$ predictive intervals for the simulator output using the heteroscedastic Gaussian process emulator, as well as the true mean and $95\%$ predictive intervals. This plot shows that the emulator indeed struggles to identify the true mean, with the true mean being much more detailed than the estimated mean. This could be considered a similar issue to the problem of choosing the degrees of freedom for smoothing splines \citep{Cantoni2002}.

To properly use an emulator as a surrogate for the simulator, this issue should be resolved, but using an excessive number of data points can be computationally expensive. In situations where additional prior knowledge of the simulator is available, a more descriptive prior mean function $m(\cdot)$ can be used, providing information that can assist in the prediction of the true mean.

In practice, sufficient knowledge of the mean function can often be lacking, which is one reason why computer experiments are conducted in the first place. However, because stochastic computer models are completely man-made, based on theoretical understanding of a real world system, modifications to the computer model are possible, and thus sometimes deterministic versions are available. This can be because a deterministic version has intentionally been made, or because the computer model was once deterministic in its development history and stochasticity was added to the computer model after its initial creation. The intuition behind the method presented below is that these deterministic versions of simulators can be informative to the overall shape of the stochastic versions' mean. After all, both the deterministic approximation and the stochastic simulator are supposed to be modelling the same real world process. In some cases, a deterministic version of a stochastic simulator can be obtained by fixing the random number generator seed, but in such cases care must be taken in determining whether the practitioner actually believes that such an approximation is actually informative to the overall mean of the stochastic simulator.

\begin{figure}[!ht]
\includegraphics[width=\textwidth]{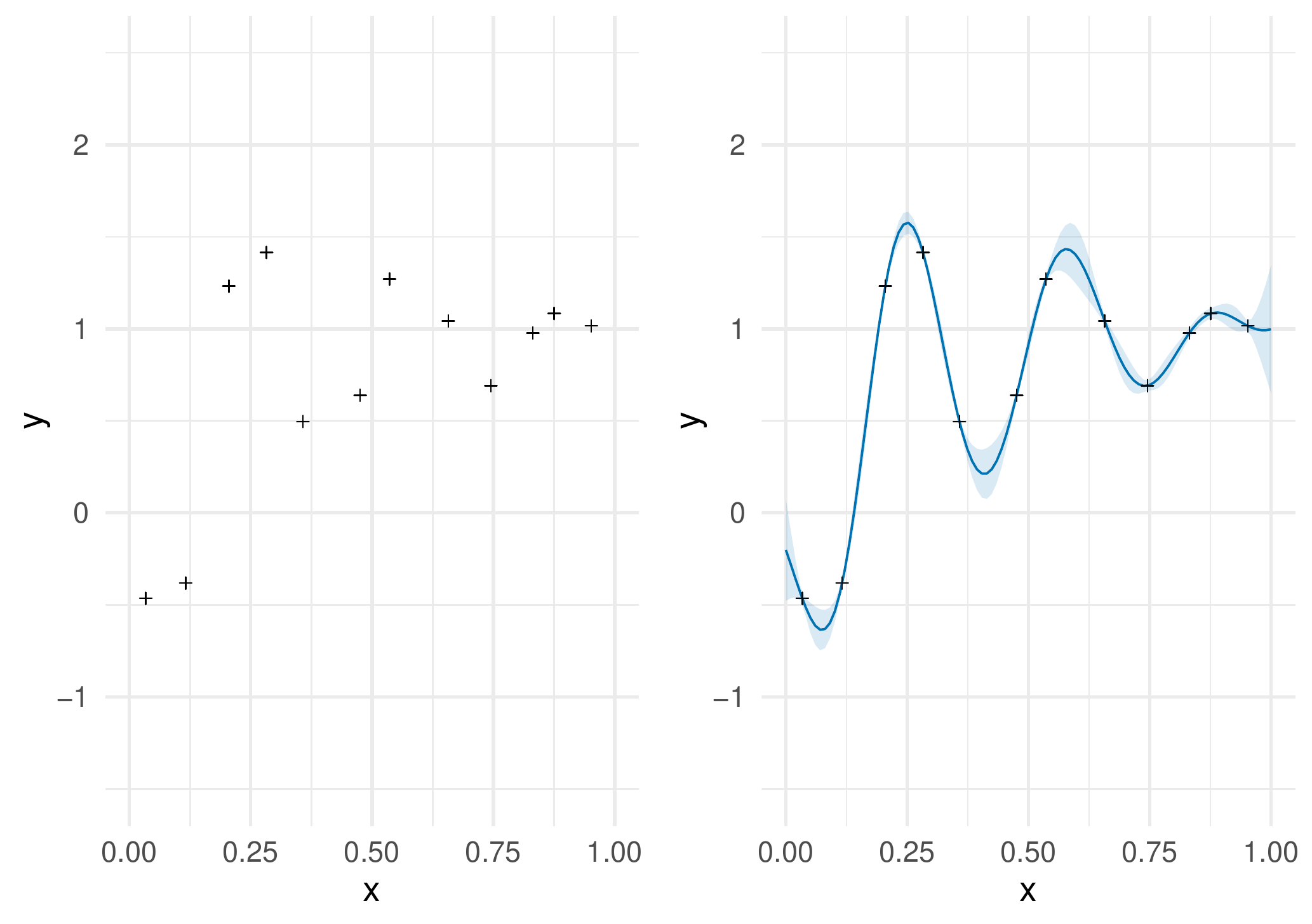}
\caption{10 Evaluations of the toy deterministic approximation simulator from equation \ref{eq:toy}, and the respective deterministic emulator predictions (right). The True mean and $95\%$ interval are superimposed in black, and the emulator mean and $95\%$ interval are in blue}
\label{fig:DetGPplot}
\end{figure}

For our toy model, we can obtain a hypothetical deterministic approximation by replacing the random component $\epsilon$ with a fixed number (in this case $\epsilon$ is replaced with 1). The plot on the left of figure \ref{fig:DetGPplot} shows 12 runs of this toy deterministic approximation simulator (chosen via a maximin latin hypercube design).

Because these runs are now samples from a deterministic simulator, we can fit a deterministic emulator to them. Using the mean function $m_{det}(\textbf{x}) =  \beta_{{det}_0} + \textbf{x}^T\bm{\beta}_{det}$; the covariance function  $K_{{det}}(\textbf{x}, \textbf{x}') = \alpha_{det}^2\prod_{i=1}^{d}\exp(-(\frac{x_i - x_i'}{l_{{det} i}})^2)$; with $\beta_{{det}_0}$ and $\bm{\beta}_{det}$ priors both $N(0,10)$; $\alpha_{det}$ having a $Inverse-Gamma(2,1)$ prior; and $l_{{det} i}$ having a $Gamma(4,4)$ prior, a deterministic Gaussian process emulator is fit to these points, and the predictions in the right of figure \ref{fig:DetGPplot} are obtained.

The shape of the emulator produced from this appears visually similar in shape to the true mean of the stochastic simulator from figure \ref{fig:HetGPplot}. This is to be expected, as the true mean of the deterministic emulator is the true mean of the stochastic simulator, offset by $(1.2-x)$.

Visually observing the predictions from this deterministic emulator would suggest that the stochastic emulator's mean should not be (approximately) linear as it is in figure \ref{fig:HetGPplot}. Nonetheless, formally incorporating this evidence remains a challenge. For the stochastic emulator, the length scale $l_i$ controls the complexity of the mean function, and so one solution could be to simply equate $l_i$ to $l_{det i}$. However, this is a wasteful use of the additional twelve simulator runs, because this only constrains the mean function to be sufficiently complex, it does not provide any specific information regarding the actual shape of the mean. 

A more holistic method might be to incorporate the entire deterministic Gaussian process into the stochastic emulator.

\section{DetHetGP} \label{sec:DetHetGP}

We have currently used three Gaussian processes to model the simulator: the Gaussian process that models the intrinsic variance $\delta^2(\cdot)$; the top level heteroscedastic Gaussian process $HetGP(\cdot)$; and the Gaussian process $DetGP(\cdot)$ that models the deterministic approximation of the simulator $\eta_{det}(\cdot)$.

We decide to model the stochastic simulator as equal to the sum of the deterministic Gaussian process and the heteroscedastic Gaussian process, equations given by equation \ref{eq:DetHetGPmodel}.

\begin{equation} \label{eq:DetHetGPmodel}
\begin{split}
	\eta(\cdot) = &DetGP(\cdot) + HetGP(\cdot)
	\\
	 & DetGP(\cdot) \sim GP(m_{det}(\cdot), K_{det}(\cdot, \cdot))
	\\
	& HetGP(\cdot) \sim GP(m(\cdot), K(\cdot, \cdot) + \delta^2(\cdot)I)
	\\
	& log(\delta^2(\cdot)) \sim GP(m_\delta(\cdot), K_\delta(\cdot, \cdot))	
\end{split}
\end{equation}

Using a sum of Gaussian processes to model a system is a common tool in deterministic emulation, having been used to include the discrepancy between the simulator and the real world \citep{Kennedy2001, Brynjarsdottir2014}; modelling non-stationary simulators \citep{Ba2012}; and modelling large-scale computer experiments \citep{Ben2011}.

Predictions for this model are (comparatively) complex. For the deterministic emulator, predictions conditional on known deterministic runs $\textbf{y}_{det} = \eta_{det}(X_{det})$ are standard \citep{Rasmussen2004} and are given in equation \ref{eq:DetGPpred}

\begin{equation}
	DetGP(X^*)|\ \textbf{y}_{det} \sim N(\mathcal{M}_{det}(X^*),\; \mathcal{V}_{det}(X^*))
\end{equation} \nonumber

\begin{equation} \label{eq:DetGPpred}
\begin{split}
	\mathcal{M}&_{det}(X^*) =\  m_{det}(X^*)\  + 
	\\
	&K_{det}(X^*, X_{det})(K_{det}(X_{det}, X_{det})^{-1}(\textbf{y}_{det} - m_{det}(X_{det})) 
	\\\\
	\mathcal{V}&_{det}(X^*) =\ K_{det}(X^*, X^*)\  - 
	\\
	&K_{det}(X^*, X_{det})(K_{det}(X_{det}, X_{det})^{-1}K_{det}(X_{det}, X^*)
	\\
\end{split}
\end{equation}

Predictions for the intrinsic (log) variance, conditional on (estimated) values at the input points $\delta^2(X)$ are also standard, and given by equation \ref{eq:DeltaGPpred}

\begin{equation}
	log(\delta^2(X^*))|\ log(\delta^2(X)) \sim N(\mathcal{M}_{\delta}(X^*),\; \mathcal{V}_{\delta}(X^*)) 
\end{equation} \nonumber

\begin{equation} \label{eq:DeltaGPpred}
\begin{split}
	\mathcal{M}&_{\delta}(X^*) = \ m_{\delta}(X^*)\  + 
	\\
	&K_\delta(X^*, X)(K_\delta(X, X))^{-1}(log(\delta^2(X)) - m_{\delta}(X))
	\\\\
	\mathcal{V}&_{\delta}(X^*) = \ K_{\delta}(X^*, X^*)\  - 
	\\
	&K_{\delta}(X^*, X)(K_{\delta}(X, X)^{-1}K_{\delta}(X, X^*)	
	\\
\end{split}
\end{equation}

Predictions for the $HetGP(\cdot)$ component are slightly more involved. This Gaussian process models the difference between the simulator and the deterministic Gaussian process (the simulator is modelled as $\eta(\cdot) = DetGP(\cdot) + HetGP(\cdot)$ and so via a simple rearrangement $HetGP(\cdot) = \eta(\cdot) - DetGP(\cdot)$), as well as including the intrinsic variance modelled by $\delta^2(\cdot)$. As such, predictions are the standard heteroscedastic predictions \citep{Binois2016} conditioned on values for $\eta(\cdot) - DetGP(\cdot)$ at the input points ( $\textbf{y} - DetGP(X)|\ \textbf{y}_{det}$ ). This leads to predictions for the heteroscedastic Gaussian process $HetGP(\cdot)$ having the form in equation \ref{eq:HetGPpred}

\begin{equation}
	HetGP(X^*)|\ \textbf{y}, \delta^2(X), \textbf{y}_{det}  \sim N(\mathcal{M}_{het}(X^*),\; \mathcal{V}_{het}(X^*))
\end{equation} \nonumber

\begin{equation} \label{eq:HetGPpred}
\begin{split}
	\mathcal{M}&_{het}(X^*) =\ m_{het}(X^*)\  +
	\\
	&K_{het}(X^*, X)((K_{het}(X, X) + \delta^2(X))^{-1}(\textbf{y} - DetGP(X)|\ \textbf{y}_{det} - m_{het}(X))
	\\\\
	\mathcal{V}&_{het}(X^*) =\ K_{het}(X^*, X^*)\  +
	\\
	&\delta^2(X^*) - K_{het}(X^*, X)((K_{het}(X, X)+\delta^2(X))^{-1}K_{het}(X, X^*)\\\\
\end{split}
\end{equation}

And so predictions for new points then become:

\begin{equation} \label{eq:DetHetGPpred}
\eta(X^*) |\ \textbf{y}, \delta^2(X), \textbf{y}_{det} = DetGP(X^*)|\ \textbf{y}_{det} + HetGP(X^*) |\ \textbf{y}, \delta^2(X), \textbf{y}_{det} 
\end{equation}

Parameter estimation for this model will also use maximum a posteriori estimates, calculated using Stan, and will have the same priors as the individual component emulators had in section \ref{sec:Intuition}. A full MCMC scheme could be used to incorporate parameter uncertainty, but this is prohibitively slow just as it is for the base heteroscedastic Gaussian process \citep{Kersting2007}.

This method takes inspiration from similar work done on emulating slow deterministic computer models using fast deterministic computer models. The use of a GP to model the simpler, more approximate computer model, and then another GP to model the discrepancy between the simple model and the more complicated model is an idea common to both this method, and the method from \cite{Kennedy2000}. 

\begin{figure}[!ht]
\includegraphics[width=\textwidth]{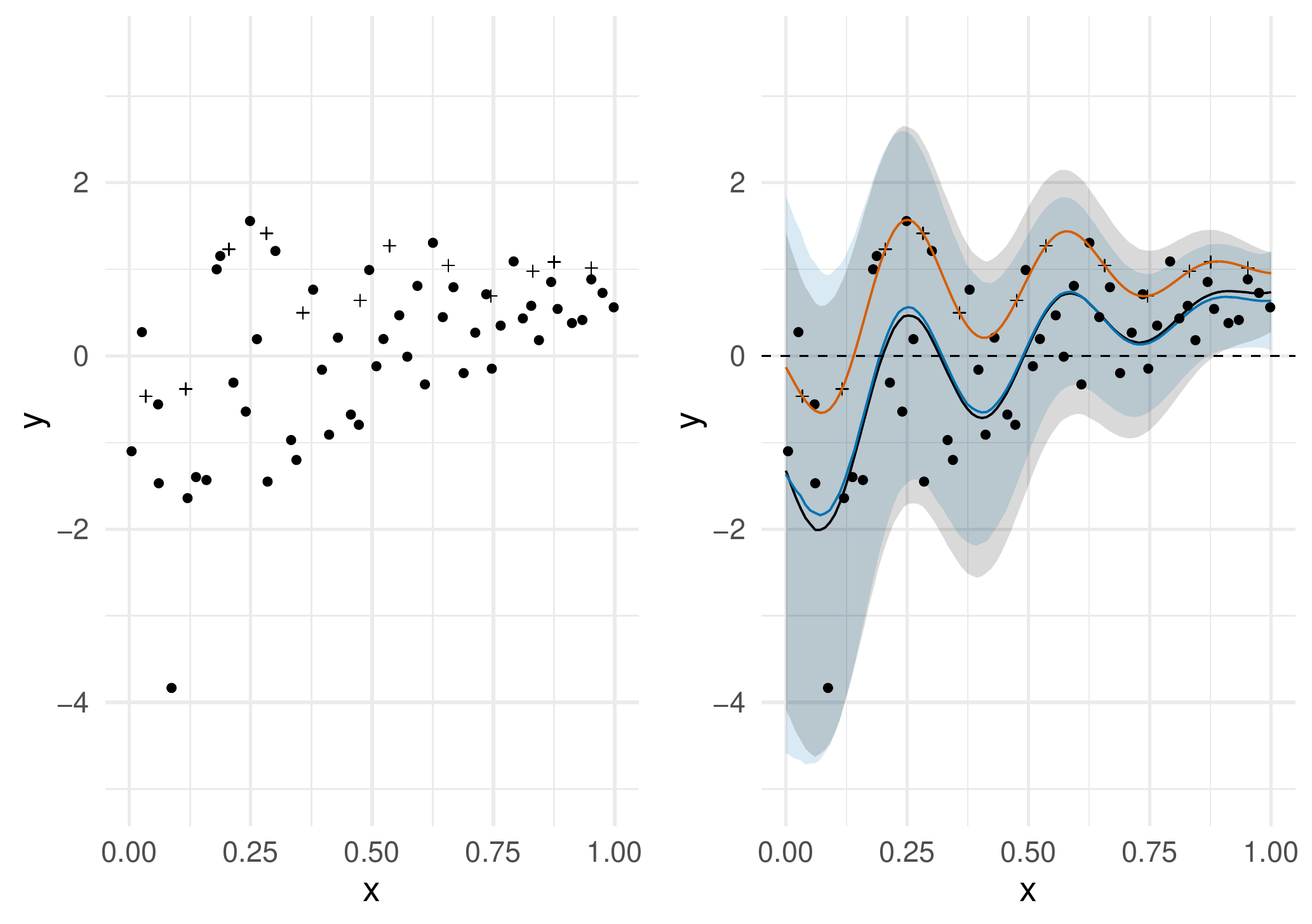}
\caption{Emulator predictions for the toy simulator, using the newly developed model that incorporates both stochastic and deterministic runs. The True mean and $95\%$ interval are superimposed in black, and the emulator mean and $95\%$ interval are in blue. The stochastic data points are circles, and the deterministic data points are plus symbols. The mean of the $DetGP$ component is in orange}
\label{fig:DetHetGPplot}
\end{figure}

Applying this model to the problem used in section \ref{sec:Intuition}, using the stochastic and deterministic data used in that section, we obtain the predictions in figure \ref{fig:DetHetGPplot}.

Clearly now the overall shape of the emulator is closer to the truth, with the appropriate periodic feature now represented in the emulator.

The orange line represents the mean of the $DetGP$ Gaussian process. Clearly, the deterministic approximation yields outputs that are overall too large compared to the stochastic mean, and thus the same is true for the $DetGP$ Gaussian process. The $DetGP$ Gaussian process can then be adjusted by the $HetGP$ Gaussian process to ensure the full emulator's mean matches with the stochastic simulator's mean. The complexity of this adjustment Gaussian process $HetGP$ is related to how good of an approximation the deterministic simulator is: in this toy example, the deterministic approximation differs from the true mean of the stochastic simulator by a constant term and a linear term, and thus the $HetGP$ Gaussian process needs to be approximately linear for the full emulator to have good fit. In this example, this appears to have happened, yielding a better fitting emulator.

\begin{figure}[!ht]
\includegraphics[width=\textwidth]{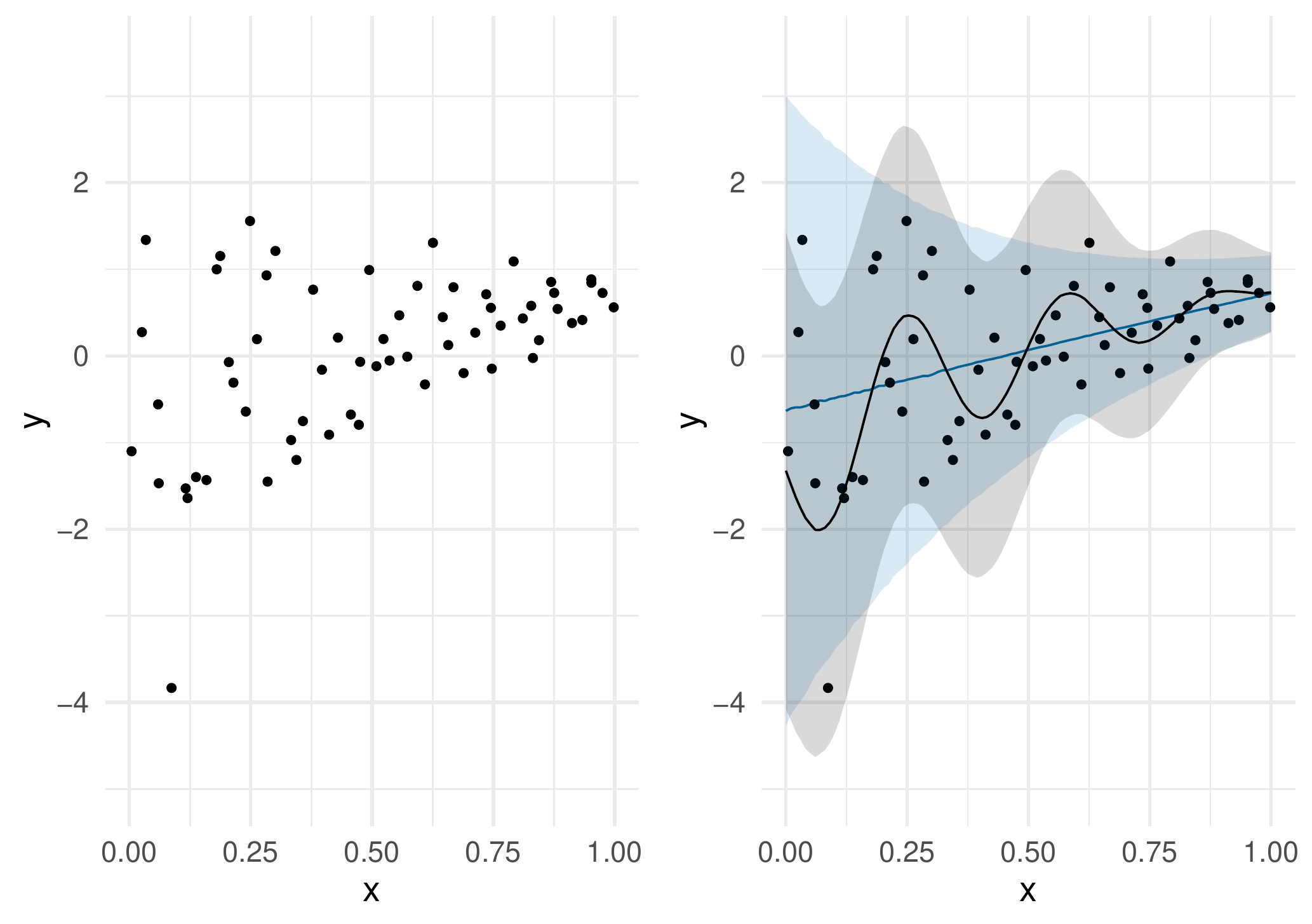}
\caption{62 Evaluations of the toy simulator from equation \ref{eq:toy} (left), and the respective heteroscedastic emulator predictions (right). The True mean and $95\%$ interval are superimposed in black, and the emulator mean and $95\%$ interval are in blue}
\label{fig:HetGPplot2}
\end{figure}

This $DetHetGP$ emulator is evidently a better surrogate for the simulator than the previous heteroscedastic emulator, but it has used an additional twelve simulator runs to do so. To show that utilising deterministic runs was an efficient use of a simulator budget, figure \ref{fig:HetGPplot2} shows the base heteroscedastic emulator fitted to the original 50 stochastic data points, plus an additional twelve stochastic data points generated from the same input x coordinates as the deterministic runs.

Here, the emulator remains substantially inferior to the emulator that uses some of the simulator budget to incorporate deterministic runs, despite using the same total number of simulator runs. This example suggests that deterministic runs can indeed be a useful tool in modelling a stochastic simulator.

\section{Performance} \label{sec:Performance}

Presented in this section are the results of several simulation experiments, comparing the performance of the newly developed method, and the default heterosecdastic Gaussian process model without deterministic runs, from here on referred to as DetHetGP and HetGP respectively.

These experiments will be conducted on quick to evaluate toy simulators, so that the predictions of these two methods can be compared to the `truth'. The purpose of these experiments is to showcase the capabilities of DetHetGP, and also show in which situations DetHetGP is most effective.

Three metrics will be used to assess the predictions of the two emulators. The first of these is the ``true mean squared error''. 100 unique prediction coordinates will be generated by a maximin latin hypercube, and the squared difference between the emulators' predictive means and the true means for these coordinates will be averaged - smaller numbers indicate better predicted mean functions. Often in the literature, ``mean squared error'' refers instead to the difference between observed values and the mean prediction of a model, which is a metric that can actually be calculated for ``real'' examples where the true mean is unknown. Here we will also generate 1000 simulator runs for each of the 100 previously generated prediction coordinates, and use these simulator runs to calculate this less accurate ``mean squared error''. Again, smaller values represent better estimated mean values.
The third metric is the scoring rule from equation 27 in \cite{Gneiting2007b} (which is also used in \cite{Binois2016} to assess stochastic emulation performance). Using the same $1000 \times 100$ simulator runs as used for the mean squared error, the total score of the predictions provides an overall measure of how good the emulator predictions are. A higher score represents better emulator predictions, with the emulator penalised for having inaccurate mean \textit{and} inaccurate variance predictions. 

Emulator performance depends on the choice of design point locations, which for these experiments will be chosen via maximin latin hypercube designs, and performance also depends on the specific observed output values of the stochastic simulator runs used to fit the emulator. Both of these are random, and thus the exact performance metric values of the two emulators will vary with each simulation. As such, the above process will be repeated 100 times for each toy simulator examined, and the lower quartiles, medians, and upper quartiles for the mean squared error and score will be reported.

To keep the performance metrics on the same scales, all the data (deterministic, stochastic, prediction) will be standardised according to the sample mean and standard deviation of the stochastic data points used to fit $HetGP$. In practice the data used for $DetHetGP$ could not be standardised according to such a data set (fewer/different stochastic data points are used to fit $DetHetGP$) but using different values to standardise the data of the two different methods would change the scales of the two sets of performance metrics, making them harder to directly compare. Regardless, the effects from such a decision should be marginal.

The first toy simulator to be examined in this way will be the one used as an example previously in sections \ref{sec:Intuition} and \ref{sec:DetHetGP}, using 60 total training points, for DetHetGP 12 of which are runs from the hypothetical deterministic approximation. We have already seen that DetHetGP appears to perform better than HetGP for this simulator, but this was for only for one set of training data points and it can also be useful to see how the improvements visually observed in section \ref{sec:DetHetGP} map to improvements in the performance metrics.

The results for this first toy simulator are given in table \ref{tab:ComplexResults}.

\begin{center}
	\begin{table} [!ht]
	\centering
    \begin{tabular}{ l  l  l  l }
    \multicolumn{4}{ c }{HetGP}\\ \hline
    & True MSE & MSE & Score \\ \hline
    Lower Quartile & 0.1787  & 0.6676 & -60990 \\ 
    Median & 0.2163 & 0.7711 & -36577 \\ 
    Upper Quartile & 0.2795  & 0.9612 & -21973 \\  \hline
	\end{tabular}

	\vspace{\baselineskip}	
	
	\begin{tabular}{ l  l  l  l }
	\multicolumn{4}{ c }{DetHetGP} \\ \hline
	& True MSE & MSE & Score \\ \hline
	Lower Quartile & 0.008292  & 0.4604 & -43764 \\ 
	Median & 0.019758 & 0.5754 & -19414 \\ 
	Upper Quartile & 0.036090 & 0.7224 &  \phantom{-0}5946 \\ \hline
    \end{tabular}
    \caption{The summary statistics of the MSE and scores of both HetGP and DetHetGP from the simulation experiment conducted on the simulator from equation \ref{eq:toy}.}
     \label{tab:ComplexResults}
    \end{table}
\end{center}

For this example, DetHetGP performs substantially better. The true mean squared error is roughly 10 times bigger for HetGP, and the score is roughly 15000 smaller. This is to be expected; the mean function is relatively complex which motivates the need for increased information about the shape of the simulator, and we have already seen how $DetHetGP$ can be useful for this toy simulator. The (non-``true'') mean squared error is much larger for both emulators, and the difference between the two is smaller (albeit still clearly present). This shows how the (non-``true'') mean squared error is a less accurate measure of mean prediction performance: improvements can still be observed, but improvements appear smaller than they really are.

The second toy simulator examined is a modified version of the one from \cite{Goldberg1997}, adapted to be 2D rather than 1D.

\begin{equation} \label{eq:Goldbergtoy}
\begin{gathered}
	\eta(x_1,x_2) = 2sin(2\pi x_1) + 2sin(2\pi x_2) + (0.5+x)\epsilon_1 + (0.5+x_2)\epsilon_2\\
	\epsilon_1 \sim N(0, 1);\quad \epsilon_2 \sim N(0, 1); 
\end{gathered}
\end{equation}

The deterministic approximation is made by replacing $\epsilon_1$ with the fixed value 0.5, and $\epsilon_2$ with the fixed value -0.5. The experiment on this simulator is run using 100 total simulation runs, 20 of which are deterministic runs for DetHetGP. Table \ref{tab:GoldbergResults} gives the respective summary statistics of the 3 performance metrics for this experiment.

\begin{center}
	\begin{table} [!ht]
	\centering
    \begin{tabular}{ l  l  l  l }
    \multicolumn{4}{ c }{HetGP}\\ \hline
    & True MSE & MSE & Score \\ \hline
    Lower Quartile & 0.04403  & 0.3804 & -24839 \\ 
    Median & 0.06231 & 0.4283 & -10441 \\ 
    Upper Quartile & 0.12049 & 0.4721 & \phantom{-00}979 \\  \hline
	\end{tabular}

	\vspace{\baselineskip}	
	
	\begin{tabular}{ l  l  l  l }
	\multicolumn{4}{ c }{DetHetGP} \\ \hline
	& True MSE & MSE & Score \\ \hline
	Lower Quartile & 0.008964  & 0.3319 & \phantom{0}-9259 \\ 
	Median & 0.013581 & 0.3646 & \phantom{-0}2512 \\ 
	Upper Quartile & 0.021878 & 0.4059 & \phantom{-}13144 \\ \hline
    \end{tabular}
    \caption{The summary statistics of the MSE and scores of both HetGP and DetHetGP from the simulation experiment conducted on the modified simulator from \cite{Goldberg1997} (equation \ref{eq:Goldbergtoy}).}
     \label{tab:GoldbergResults}
    \end{table}
\end{center}

Once again, DetHetGP provides substantial improvements over HetGP. The improvements are slightly less than before, which could be because the mean function is simpler for this example. Regardless, this example provides some evidence that the utility of $DetHetGP$ was not just a quirk of the previous example, and that $DetHetGP$ can still yield improvements for a non-1D simulator. Again, the substantial improvements in the mean predictions, which can be clearly seen from the ``true'' mean squared error, are less pronounced for the ``non-true'' mean squared errors. This provides further evidence that changes in the ``non-true'' mean squared errors undersell the true changes in the mean predictions. This effect is present in the remaining toy simulation experiments, but will no longer be commented on.

The next simulator considered is the one from \cite{Binois2017}, also given in equation \ref{eq:Binoistoy}. The deterministic approximation for this simulator is made by fixing $\epsilon$ as 1. This then leads to a poorer deterministic approximation than the previous two examples, as now the deterministic approximation differs from the stochastic simulators mean by $1.1 + sin(2\pi x)$, which is much more complicated than the previous linear differences. This will then test how well $DetHetGP$ performs when the deterministic approximation is less informative of the stochastic simulator's mean. Using 60 total data points, 15 of which are deterministic for $DetHetGP$ yields the results in table \ref{tab:BinoisResults}.

\begin{equation} \label{eq:Binoistoy}
\begin{gathered}
	\eta(x) = (6x - 2)^2sin(12x - 4) + (1.1 + sin(2\pi x))\epsilon\\
	\epsilon \sim N(0, 1)
\end{gathered}
\end{equation}

\begin{center}
	\begin{table} [!ht]
	\centering
    \begin{tabular}{ l  l  l  l }
	\multicolumn{4}{ c }{HetGP} \\ \hline
	& True MSE & MSE & Score \\ \hline
	Lower Quartile & 0.007932  & 0.08527 & 154488 \\ 
	Median & 0.012922 & 0.09165 & 168005 \\ 
	Upper Quartile & 0.019838 & 0.09796 & 179477 \\ \hline
	\end{tabular}

	\vspace{\baselineskip}	
	
	\begin{tabular}{ l  l  l  l }
	\multicolumn{4}{ c }{DetHetGP}\\ \hline
    & True MSE & MSE & Score \\ \hline
    Lower Quartile & 0.0039057  & 0.08061 & 160791 \\ 
    Median & 0.0061066 & 0.08428 & 172852 \\ 
    Upper Quartile & 0.0094305 & 0.08915  & 181012\\  \hline
    \end{tabular}
    \caption{The summary statistics of the MSE and scores of both HetGP and DetHetGP from the simulation experiment conducted on the Binois simulator (equation \ref{eq:Binoistoy}).}
     \label{tab:BinoisResults}
    \end{table}
\end{center}

For this toy simulator, $DetHetGP$ performs only marginally better than $HetGP$: The score is only larger by a few hundred. However, the ``true'' mean squared error is half the size, so despite the decreased informativeness of the deterministic approximation, the mean prediction of $DetHetGP$ is still substantially  better than that of $HetGP$. As the deterministic approximation becomes less informative, the utility of $DetHetGP$ decreases. This is an obvious effect; if a simulator is not informative about a system (in this case, if a deterministic simulator is not informative about a stochastic simulator), then information from said simulator will of course be poor.

We have given a few examples showing $DetHetGP$ can provide significant improvements over $HetGP$. The following examples will then highlight some of the key limitations of $DetHetGP$, which serves to identify when it will be a useful method, and when it will not be.  We have already briefly discussed one limitation, which is if the deterministic approximation is a poor approximation, then the utility $DetHetGP$ provides is decreased.

The first example will highlight how too few deterministic data points can limit the utility of $DetHetGP$, and in extreme cases make it worse than $HetGP$. The same toy simulator from sections \ref{sec:Intuition} and \ref{sec:DetHetGP} will be used, but this time with 200 total data points (ensuring that $HetGP$ should perform better than seen previously), but with only 3 assigned as deterministic points for $DetHetGP$. The results of the simulation experiment are given in table \ref{tab:TooFewNdet}

\begin{center}
	\begin{table} [!ht]
	\centering
    \begin{tabular}{ l  l  l  l }
    \multicolumn{4}{ c }{HetGP}\\ \hline
    & True MSE & MSE & Score \\ \hline
    Lower Quartile & 0.02591  & 0.5372 & -26503 \\ 
    Median & 0.06393 & 0.6020 & -14037 \\ 
    Upper Quartile & 0.16676  & 0.7051 & \phantom{0}-4056 \\  \hline
	\end{tabular}

	\vspace{\baselineskip}	
	
	\begin{tabular}{ l  l  l  l }
	\multicolumn{4}{ c }{DetHetGP} \\ \hline
	& True MSE & MSE & Score \\ \hline
	Lower Quartile & 0.11297  & 0.5938 & -32853 \\ 
	Median & 0.14233 & 0.6704 & -21859 \\ 
	Upper Quartile & 0.16977 & 0.7481 & -13807 \\ \hline
    \end{tabular}
    \caption{The summary statistics of the MSE and scores of both HetGP and DetHetGP from the simulation experiment conducted on the toy simulator from equation \ref{eq:toy} using 200 total data points, only 3 of which are deterministic data points for $DetHetGP$.}
     \label{tab:TooFewNdet}
    \end{table}
\end{center}

Here the score and mean squared error are both significantly worse for $DetHetGP$. With so few deterministic data points, the deterministic emulator cannot have good fit. Not only does this then waste 3 simulator runs that could have been stochastic runs instead, but the resulting poorly fit deterministic emulator then actually \emph{misinforms} the overall emulator, leading to poor predictions.

The second limitation is for the opposite problem: too many deterministic runs. Using the same toy example as above, this time with 50 total data points, 35 of which are deterministic for $DetHetGP$, gives the results in table \ref{tab:TooManyNdet}.

\begin{center}
	\begin{table} [!ht]
	\centering
    \begin{tabular}{ l  l  l  l }
    \multicolumn{4}{ c }{HetGP}\\ \hline
    & True MSE & MSE & Score \\ \hline
    Lower Quartile & 0.1787  & 0.6676 & \phantom{0}-60990 \\ 
    Median & 0.2163 & 0.7711 & \phantom{0}-36577 \\ 
    Upper Quartile & 0.2795  & 0.9612 & \phantom{0}-21973 \\  \hline
	\end{tabular}

	\vspace{\baselineskip}	
	
	\begin{tabular}{ l  l  l  l }
	\multicolumn{4}{ c }{DetHetGP} \\ \hline
	& True MSE & MSE & Score \\ \hline
	Lower Quartile & 0.016289  & 0.5088 & -107327 \\ 
	Median & 0.050736 & 0.6695 & \phantom{0}-54675 \\ 
	Upper Quartile & 0.122364 & 0.8867 & \phantom{0}-15398 \\ \hline
    \end{tabular}
    \caption{The summary statistics of the MSE and scores of both HetGP and DetHetGP from the simulation experiment conducted on the toy simulator from equation \ref{eq:toy} using 50 total data points, 35 of which are deterministic data points for $DetHetGP$.}
     \label{tab:TooManyNdet}
    \end{table}
\end{center}

Here the score is much worse for $DetHetGP$. This is because so much of the simulation budget is used on deterministic runs, that there is not enough stochastic data points to estimate the stochastic simulator's variance. In this example however, the mean squared error is still considerably better for $DetHetGP$, the excessive number of deterministic points has lead to a good estimation of the stochastic simulators mean. In other cases, too few stochastic data points could also lead to a poor fit of the adjustment Gaussian process, yielding poorer predictions for \emph{both} the mean and variance. Evidence of this can be observed by noting that the mean squared error using 35 deterministic points is higher than that from table \ref{tab:ComplexResults} where only 12 deterministic points were used.

\section{Examples} \label{sec:Examples}
In this section, the method will be applied to two, more realistic, examples. 

The first of these is a basic Suceptible-Infected-Recovered (SIR) model using the  Individual Contact Model (ICM) from the EpiModel package outlined in \cite{Jenness2018}. This stochastically models a population, where individuals can be: susceptible to some disease; currently infected with the disease; or recovered from (and now immune to) the disease. This is still a fairly quick to evaluate, overly simplistic simulator, but it serves as a more authentic example than the toy simulators from the previous section. We have chosen the parameters 0.01 for the rate of infection risk interactions between individuals, an initial infected rate of 5 out of a population of 1000, and we shall record the infected proportion 300 time steps later. The two parameters we shall vary as our inputs for the 2 emulators are: the probability of infection which will be allowed to vary between 0.5 and 1, and the recovery rate which will vary between 0 and 0.01.
For the deterministic approximation, the package also includes a  Deterministic Compartmental Model (DCM) which takes the same inputs and yields the same outputs.

To compare the performance of $HetGP$ and $DetHetGP$, we no longer have access to the ``true mean" of the stochastic simulator to calculate the ``true mean squared error''. The simulator is also too slow to easily allow $100 \times 1000 \times 100$ simulator runs to obtain values for the score and mean squared error, as was done before. Instead, for this example, 200 simulator runs will be obtained for coordinates chosen by a latin hypercube design, and the score and mean squared error evaluated for both $HetGP$ and $DetHetGP$. This will be repeated 100 times, with the lower quartiles, medians and upper quartiles reported, as before. 100 total simulator runs will be used to fit the emulators, with 20 assigned as deterministic points for $DetHetGP$. Table \ref{tab:ABM} gives these values.

\begin{center}
	\begin{table} [!ht]
	\centering
    \begin{tabular}{ l  l  l  l }
	\multicolumn{3}{ c }{HetGP} \\ \hline
	& MSE & Score \\ \hline
	Lower Quartile & 0.3536 & 103.69 \\ 
	Median & 0.4414 & 142.26 \\ 
	Upper Quartile & 0.5549 & 177.68 \\ \hline
	\end{tabular}

	\vspace{\baselineskip}	
	
	\begin{tabular}{ l  l  l  l }
	\multicolumn{3}{ c }{DetHetGP}\\ \hline
    &  MSE & Score \\ \hline
    Lower Quartile & 0.2921 & 110.007 \\ 
    Median & 0.3407 & 146.575 \\ 
    Upper Quartile & 0.4184  & 184.502\\  \hline
    \end{tabular}
    \caption{The summary statistics of the score for both HetGP and DetHetGP from the simulation experiment conducted on the SIR simulator.}
     \label{tab:ABM}
    \end{table}
\end{center}

Here we can see that the score is marginally improved for $DetHetGP$, and the mean squared error is substantially smaller, with the median for $DetHetGP$ even being smaller than the lower quartile for $HetGP$. Not only does $DetHetGP$ appear to be the preferred emulator here, but this example is one where the deterministic approximation is actually much faster than the stochastic version due to it's increased simplicity. For a hundred input points (chosen by a maximin latin hypercube), the stochastic simulator takes 13.89 seconds to run, on the other hand the deterministic approximation takes only 3.08 seconds to run. For the SIR simulator, the difference in run time increases as the total population number for the system increases, and the number of time steps simulated increases: for a total population of 10000, run for 1000 time steps, but all the other parameters kept the same as before, the stochastic simulator then takes 101.50 seconds to run and the deterministic approximation takes 9.38 seconds to run.

\begin{figure}[!ht]
\includegraphics[width=\textwidth]{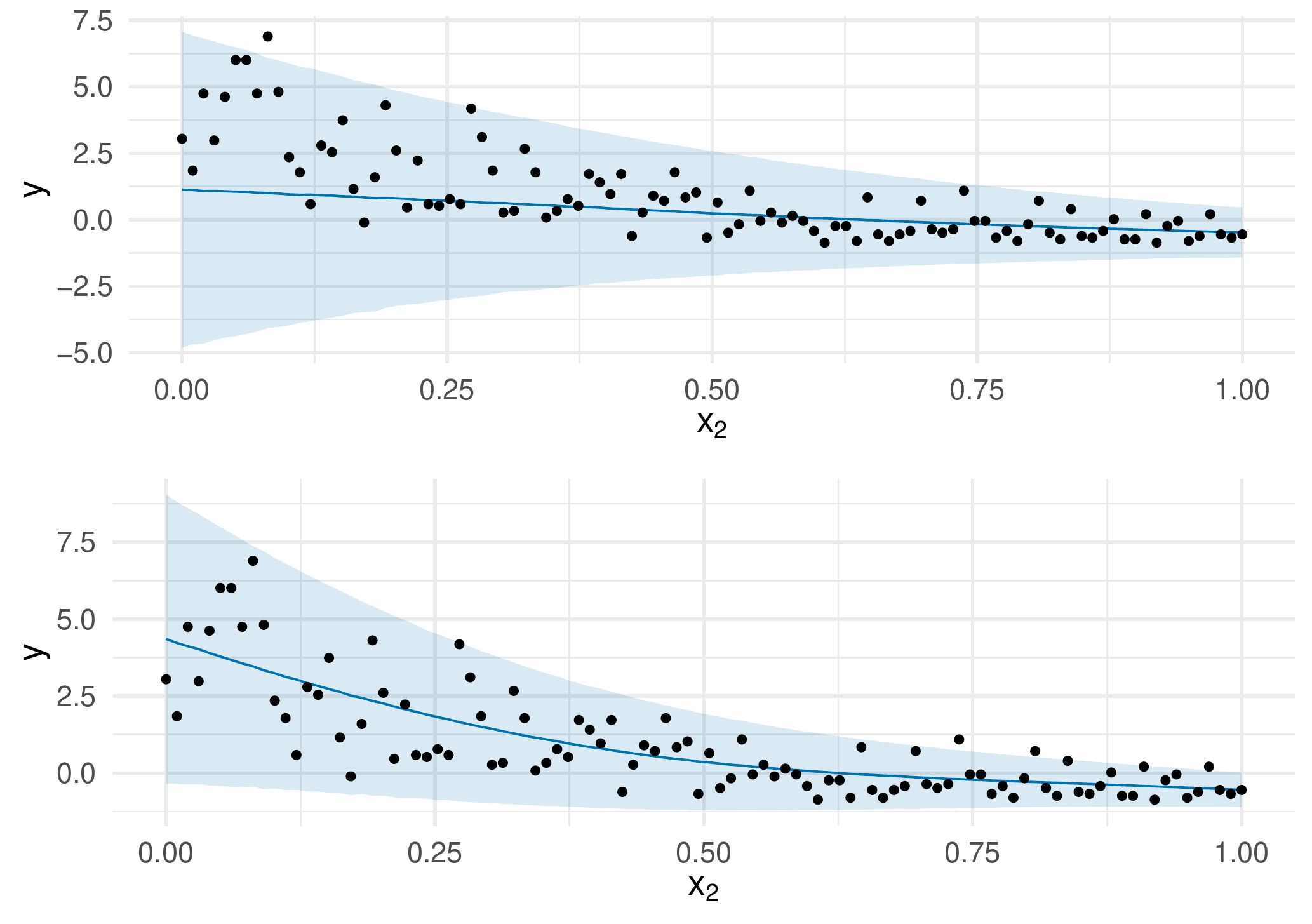}
\caption{Emulator predictions for both $HetGP$ and $DetHetGP$. The predictions are for simulator runs where the infection rate is constant at 1, and only the recovery rate varies. Also superimposed are 100 additional simulator runs where the infection rate is kept constant at 1.}
\label{fig:ABMCrossSectionPlot}
\end{figure}

To visually show the improvements $DetHetGP$ yields, using one set of training simulator runs from the previously described designs, figure \ref{fig:ABMCrossSectionPlot} shows predictions from both emulators with the infection rate $x_1$ kept constant at 1, and only the recovery rate is varied. Superimposed on this plot is also 100 out of sample prediction simulator runs where the infection rate was also kept fixed at 1. The mean predictions from $HetGP$ are approximately linear, missing the sharper increase for lower values of the recovery rate, and instead a larger variance is predicted for these values - large observed simulator runs were probably interpreted by the emulator as being because of a larger variance rather than because of a larger mean. $DetHetGP$ on the other hand does estimate the sharp increase in the mean function, and does not feature an overly large variance estimate for low values of the recovery rate.  Other cross sections of the emulator's prediction surface could have been presented, and a similar pattern exists for if the recovery rate is fixed at 0 and the infection rate is allowed to vary, but this cross-section is presented as one potential explanation for the perceived improved accuracy of $DetHetGP$ for this simulator. Also note from the plot that the variance for $DetHetGP$ seems smaller than that of $HetGP$ for larger values of the recovery rate. This difference is marginal, but it does lead to some of the out-of-sample points lying outside the $95\%$ interval in this region of space. Perhaps this is why the observed improvements in score are less than the improvements in mean squared error - the substantially improved mean function at lower values of the recovery rate comes at a cost of a slightly worse variance function for high values of the recovery rate (where the mean is much more simple).

As a second, more real-world example, we apply $DetHetGP$ and $HetGP$ to a simulator for modelling the energy usage of a building \citep{Crawley2000}. The modelled building is a hospital, taken from a reference hospital file \citep{Deru2011}, and the input variables $\textbf{x}$ considered are: wall concrete thickness, wall insulation thickness, roof insulation thickness, floor concrete thickness, and window size (as a percentage of the total wall height).

Typically, a single year of weather data is input, and the simulator outputs the energy usage the building would have, given that year of weather will be observed. Work has been done creating example weather files that best represent a typical year of weather \citep{Eames2015}. Instead, in this article, every time the building simulator ran, a new weather file is randomly generated by combining sampled weeks from historical records, stratified according to season. This then makes the building simulator stochastic, with the output being a random draw for what the energy usage could be, accounting for the fact that in reality the weather is effectively random and we do not know what the weather will be. A deterministic approximation of this stochastic simulator is then easily available as simply the same simulator but with the weather fixed again.

This simulator is then even more expensive than the SIR simulator, and even $100 \times 200$ runs become infeasible to score the two methods. Instead only one set of 500 scoring points will be generated, with input values chosen by a maximin latin hypercube, and the scores and mean squared errors for each method reported. 
For this simulator, 200 data points are used to fit the two emulators, with each simulator run taking a bit more than a minute. For $DetHetGP$ 50 of the data points will be deterministic runs, coordinates chosen by a maximin latin hypercube, and 150 will be stochastic also chosen via a maximin latin hypercube design. For $HetGP$ the same stochastic runs will be used, as well as an additional 50 stochastic points with coordinates the same as those for the deterministic points.
In this example, the data sets will be standardised according to the sample mean and standard deviation of the shared 150 stochastic data points.

$DetHetGP$ receives a score of -145.72 and an MSE of 0.5460, whereas $HetGP$ receives a score of -154.35 and an MSE of 0.6138. Here $DetHetGP$ seems to perform better than $HetGP$ on the building model. With only one set of 500 data points used to obtain this score, it does become more difficult to assess whether this improvement is ``real'' or just an artefact of this specific example. 

\begin{figure}[!ht]
\centering
\includegraphics[width = \textwidth]{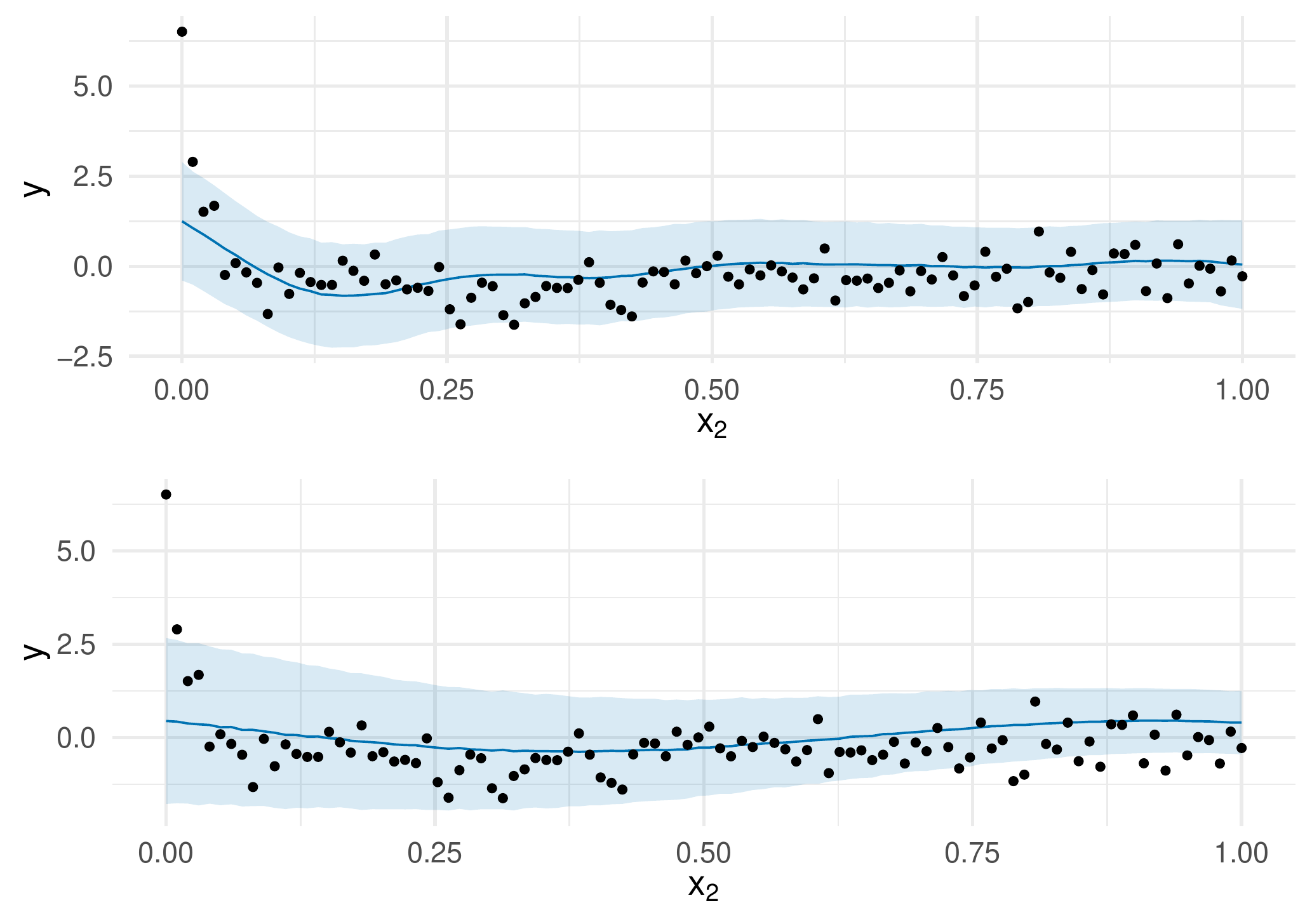}
\caption{Emulator predictions for second input of the building model (wall insulation thickness), when all other inputs are kept fixed at 0.5. $DetHetGP$ predictions are plotted above, and $HetGP$ predictions are below. Also superimposed are 100 additional simulator runs where the the remaining 4 inputs were kept constant at 0.5.}
\label{fig:2BuildingPlots}
\end{figure}

To further investigate this perceived difference, figure \ref{fig:2BuildingPlots} shows the emulator predictive distribution for wall insulation thickness, keeping all other inputs fixed (at 0.5). 100 additional out-of-sample simulator runs are also obtained, where all other input points are also fixed at 0.5.

The predictions from $DetHetGP$ are characterised by a sharp decrease in energy usage for very low values of wall insulation, after which improvements in energy efficiency seem to stabilise. This is consistent with our experience of the building model. $HetGP$ does not yield this characteristic, instead it is more defined by a significant decrease in the variability of the energy usage as wall insulation thickness increases. If we were to assume that the trend discovered by $DetHetGP$ is true, then it is likely that extremely large observed values of energy usage were instead assumed by $HetGP$ to be the result of a increased variance, rather than a sharp increase in the mean. The superimposed out-of-sample data points appear to agree more so with the predictions from $DetHetGP$, suggesting this emulator is indeed better. An important remark the data points point to is that the sharp increase for low values of wall insulation thickness, that is predicted by $DetHetGP$ but not by $HetGP$, appears to actually be even sharper than predicted by $DetHetGP$. It stands to reason that both emulators are thus poorly modelling the mean, with $DetHetGP$ being the better of the two. This issue suggests that the number of deterministic points is too low. With more deterministic data points, especially for simulator runs for low values of wall insulation thickness, the mean prediction for $DetHetGP$ should improve.

It isn't certain, but it seems likely that $DetHetGP$ performs better than $HetGP$. Including deterministic runs leads to a more accurately shaped mean function, and because the two emulators disagree with the mean function shape, it is probable that $HetGP$ has estimated the mean function poorly. 

\section{Conclusions} \label{sec:Conclusions}
We have presented a method for including deterministic runs in the emulation of stochastic simulation experiments. By utilising a deterministic approximation of the stochastic simulator, a less noisy view into the general shape of the mean function can be learnt.
Including deterministic runs can produce better fitting emulators, especially when the mean function is complex, and sufficient prior knowledge of the mean function is lacking, or the simulation budget is insufficiently large.

$HetGP$ is not the only method for modelling a stochastic simulator with a Gaussian process. The variance could instead be modelled with a simple parametric form, as by \cite{Boukouvalas2014}. Similarly, homoscedasicity may be an assumption a practitioner is willing to make, and thus a fixed variance, as in basic Gaussian process regression \citep{Rasmussen2004}, may be viable. It would be interesting to see if $DetHetGP$ remains useful in these cases, and whether it performs better in such scenarios. Similarly, other methods apart from Gaussian processes exist for flexibly modelling heteroscedastic systems, such as the method by \cite{Pratola2017}. It would also be interesting to see if deterministic approximations can be incorporated into other methods.

Additional work remains on how to decide on the number of deterministic runs to include, too few and the deterministic Gaussian process will not accurately represent the shape of the mean; too many and not enough stochastic data points will be generated, yielding a poor estimate of the variance. This is the main limitation of $DetHetGP$. One potential idea would be to first fit a deterministic emulator to the deterministic approximation, using as many deterministic points as required to build an adequate emulator (this could be assesed by leave-one-out validation, or diagnostic methods reliant on out of sample validation points such as those from \cite{Bastos2009}). Then, once this is done, the remainder of the simulation budget could be assigned as stochastic data points. Nevertheless, the design of the simulation experiment remains an open question.

Another, potentially promising, idea for design would be the inclusion of replicated simulator runs in the fitting of the emulator. Replicates are often used in the fitting of heteroscedastic emulators \citep{Ankenman2010, Boukouvalas2014a, Binois2017}, with the goal of obtaining a better understanding of the simulator's mean and variance for the observed input points. With $DetHetGP$ substantially improving the mean of stochastic emulator predictions, at the expense of fewer stochastic simulator runs, and sometimes at the expense of a worse variance prediction, it would be interesting to see whether combining replicates with incorporating deterministic approximations would yield an improved emulator overall.

Another extension to better estimate the variance process, as well as the mean process, might involve extracting yet more information from the deterministic simulator runs. One might assume that the distance of the deterministic points to the stochastic points is also informative about the variance of the system. As an intuitive example, if the stochastic simulator has very little variability and is almost deterministic, one would be surprised to find the deterministic approximation to be very different to the stochastic simulator's mean - the addition of a very mild amount of stochasticity to the deterministic simulator should probably not significantly change the mean. On the other hand, if the amount of variance is very large, then it is not unreasonable to imagine the mean of the deterministic simulator might be very different to the mean of the stochastic simulator - a large degree of added stochasticity suggests a large degree of scientific uncertainty about the true process, and thus an increased chance that the deterministic approximation would not `sync up'' with the stochastic simulators mean. If such a belief exists, then the spread of the variance of the stochastic simulator might be better estimated by also comparing the stochastic simulator runs to the deterministic computer runs  as well as just using the spread of the stochastic runs.
\clearpage

\bibliography{library}

\appendices
\section{DetHetGPprior}\label{sec:DetHetGPprior}
All the emulators fit in this article have had an additional `nugget variance' added for computational reasons \cite{Neal1997}. A value of 1e-4 was required for this nugget variance to prevent computational issues in inverting the covariance matrices. 

\begin{figure}[!ht]
\includegraphics[width=\textwidth]{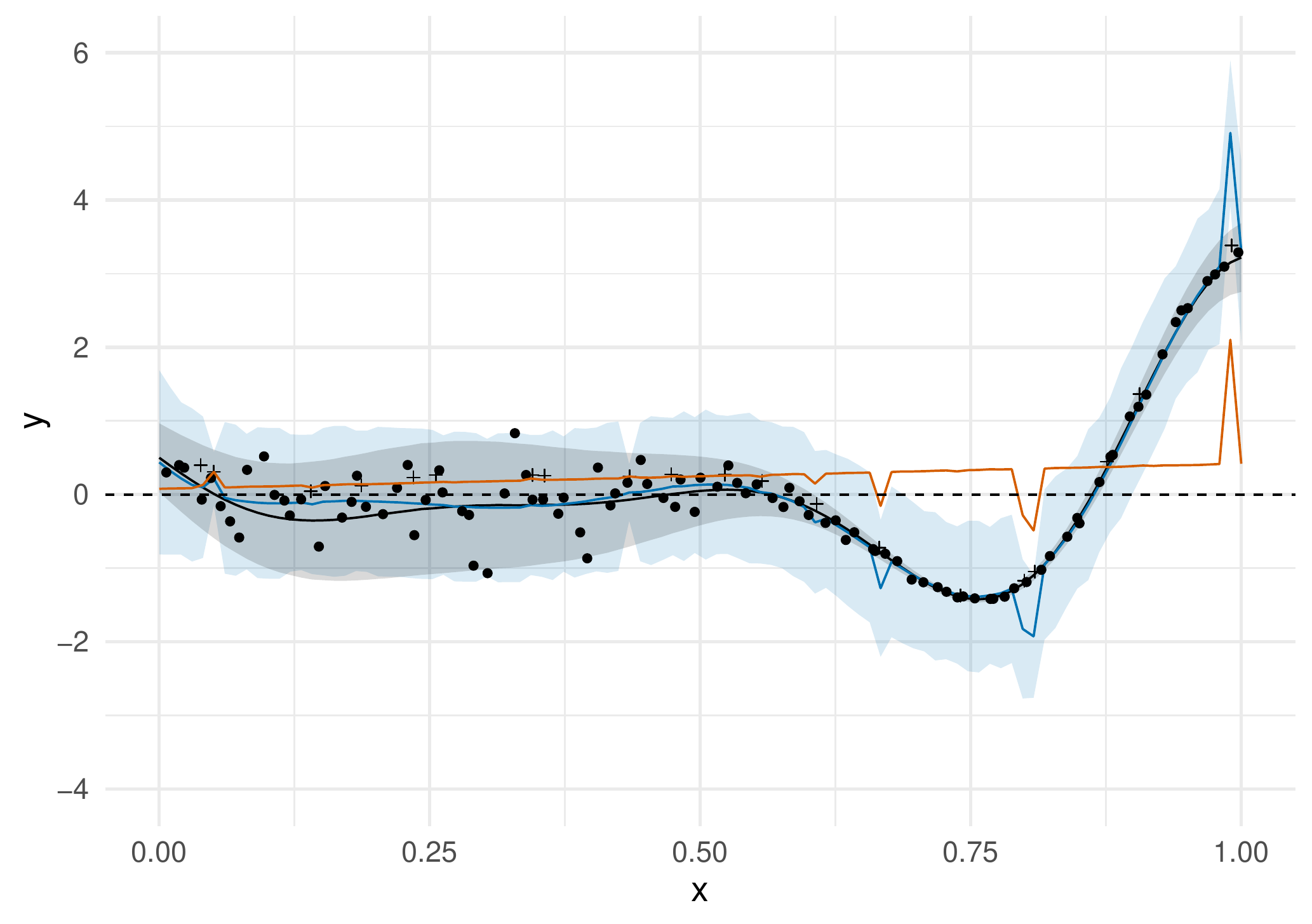}
\caption{Emulator predictions for the toy simulator from \cite{Binois2017}, using the newly developed model that incorporates both stochastic and deterministic runs. The True mean and $95\%$ interval are superimposed in black, and the emulator mean and $95\%$ interval are in blue. The stochastic data points are circles, and the deterministic data points are plus symbols. The mean of the $DetGP$ component is in orange.}
\label{fig:smallLplot}
\end{figure}

For the stochastic emulator that incorporates deterministic runs, the $l_{det i}$ parameter has been given a slightly different prior than it does for the simpler deterministic emulator. instead $l_{det i} = 0.05 + l^*_{det i}$, and $l^*_{det i}$ has a $Gamma(4,4)$ prior. This constrains $l_{det i}$ to be greater than 0.05. Figure \ref{fig:smallLplot} gives an example of what can happen in practice if $l_{det i}$ is not explicitly constrained to be larger than zero.

$l_{det i}$ has an estimated value of 0.00150 here, which is very small. This leads the deterministic Gaussian process (plotted in orange) to be approximately a straight line, with steep jumps towards the observed deterministic points (which is an established problem discussed by \cite{Andrianakis2012}, caused by the inclusion of a nugget variance). This leads to the final stochastic emulator also having steep unnecessary jumps because there is not enough stochastic data for the $HetGP$ to smooth them out. Additionally, the deterministic emulator being approximately linear also leads to extraneous variance in the deterministic emulator predictions. This additional variance then also mostly accounts for the intrinsic variance of the stochastic simulator, leading to smaller estimates for $\delta^2(X^*)$, and a less flexible estimated variance process.

This issue could be prevented by decreasing the value of the nugget variance for the deterministic Gaussian process, but that expectedly leads to computational errors. An alternative solution is the one implemented, fixing $l_{det i}$ to be sufficiently larger than zero.

\end{document}